\documentclass[final,5p,times,twocolumn,authoryear]{elsarticle}
\usepackage{caption}

\usepackage{amssymb}
\usepackage{lipsum}
\usepackage{amsmath}
\usepackage{amsthm}
\usepackage{natbib}

\usepackage{booktabs} 
\usepackage{algorithm} 
\usepackage{algpseudocode} 
\usepackage{array}
\newtheorem{theorem}{Theorem}
\newtheorem{definition}{Definition} 
\newtheorem{lemma}{Lemma}

\usepackage{multirow}

\journal{knowledge-based systems}

\begin{document}

\begin{frontmatter}

\title{Prompt-Matcher:  Leveraging Large Models to Reduce Uncertainty in Schema Matching Results}


\author[1]{Longyu Feng}
\ead{fly_fenglongyu@outlook.com}

\author[1]{Li Huahang}
\ead{huahang.li@polyu.edu.hk}

\author[1]{Chen Jason Zhang}

\ead{jason-c.zhang@polyu.edu.hk}
\affiliation[1]{
            organization={The Hong Kong Polytechnic University, },
            city={Hong kong},
            country={China}}

\begin{abstract}
Schema matching is the process of identifying correspondences between the elements of two given schemata, essential for database management systems, data integration, and data warehousing.  For datasets across different scenarios, the optimal schema matching algorithm is different.  For single algorithm, hyperparameter tuning also cases multiple results.  All results assigned equal probabilities are stored in probabilistic databases to facilitate uncertainty management.  The substantial degree of uncertainty diminishes the efficiency and reliability of data processing, thereby precluding the provision of more accurate information for decision-makers.  
To address this problem, we introduce a new approach based on fine-grained correspondence verification with specific prompt of Large Language Model.

Our approach is an iterative loop that consists of three main components:
(1) the correspondence selection algorithm,
(2) correspondence verification, and
(3) the update of probability distribution.  
The core idea is that correspondences intersect across multiple results, thereby linking the verification of correspondences to the reduction of uncertainty in candidate results. 
 The task of selecting an optimal correspondence set to maximize the anticipated uncertainty reduction within a fixed budgetary framework is established as an NP-hard problem.  We propose a novel $(1-1/e)$-approximation algorithm that significantly outperforms brute algorithm in terms of computational efficiency.  To enhance correspondence verification, we have developed two prompt templates that enable GPT-4 to achieve state-of-the-art performance across two established benchmark datasets.  Our comprehensive experimental evaluation demonstrates the superior effectiveness and robustness of the proposed approach.

\end{abstract}




\end{frontmatter}




\section{Introduction}
\label{introduction}

\subsection{Background and Motivation}
Schema matching is crucial for identifying equivalent elements across diverse schemata in heterogeneous data repositories. This technique underpins numerous applications, including database management systems, data migration, warehousing, mining, and knowledge discovery.  The rise of data science has further intensified the need for advanced data integration and management solutions.

\begin{table}
\centering
\caption{Example of Candidate Result Set}
\label{tab: Example of possible matchings result}
\resizebox{\columnwidth}{!}{\begin{tabular}{ll}
  \hline
  Candidate Result & Confidence \\ 
  \hline
  $s_1$=(Employee, EmployeeInfo):[(ID, EmployeeID),& \multirow{4}{*}{0.55}\\
  (Name, (First Name, Last Name)),\\
  (Email, Email Address), (Address, Home Address),\\
  (Gender, Sex)]  \\
  \hline
  $s_2$=(Employee, EmployeeInfo):[(ID, EmployeeID),& \multirow{4}{*}{0.25}\\
  ,(Age, Years of Experience),(Email, Email Address)\\
  (Address, Home Address)m(Gender, Sex)]  \\
  \hline
  $s_3$=(Employee, EmployeeInfo):[(ID,EmployeeID),& \multirow{3}{*}{0.20}\\
  (Name, (First Name, Last Name )),(Email, Email Address),\\
  (Gender, Sex)]  \\ 
  \hline
  Correspondences & Probability \\
  \hline
  $c_1$= [ID, EmployeeID] & 1.0 \\
  $c_2$= [Name, [First Name, Last Name ]] & 0.80 \\
  $c_3$= [Email, Email Address] & 0.75 \\
  $c_4$= [Address, Home Address] & 0.80 \\
  $c_5$= [Age, Years of Experience] & 0.25 \\
  $c_6$= [Gender, Sex] & 1.0 \\
  \hline
\end{tabular}}
\end{table}  

\begin{figure*}[!ht]
  \centering
  \includegraphics[width=\linewidth,height=0.25\linewidth]{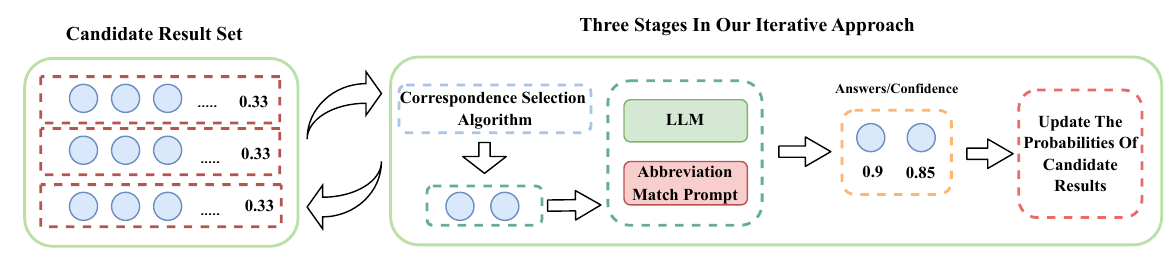}
  \caption{Prompt-Matcher: The total budget is split the budget into k budget shares.  At each iteration, correspondence selection algorithm try to selection the correspondence subset that maximize the expectation of uncertainty reduction.  Then, LLM proposes the answers and their confidence of correspondence verification.  Finally, the probabilities of candidate results are updated with the answers and the confidences.}
  \label{fig: process of prompt-matcher}
\end{figure*}

Over the years, a substantial number of semi-automated and fully automated schema matching algorithms have been developed and extensively investigated \cite{Bonifati_Velegrakis_2011,bernstein2011generic,rahm2001survey,Shvaiko_Euzenat_2005}.  Schema matching tools exhibit diverse focus areas, encompassing linguistic, structural, and instance-based features.  While these tools demonstrate satisfactory performance on specific datasets, each schema matching approach presents distinct strengths and limitations. No single schema matching methodology has been established to consistently outperform alternative approaches across various scenarios\cite{Koutras_Siachamis_Ionescu_Psarakis_Brons_Fragkoulis_Lofi_Bonifati_Katsifodimos_2021}.  Most schema matching algorithms depend on complex parameter configurations to achieve optimal performance; however, in real-world applications, their effectiveness is often compromised due to the practical challenges of maintaining precise parameter tuning, resulting in suboptimal matching accuracy\cite{Koutras_Siachamis_Ionescu_Psarakis_Brons_Fragkoulis_Lofi_Bonifati_Katsifodimos_2021}.  Furthermore, certain schema matching tools generate a set of candidate results for users\cite{Dong_Halevy_Yu_2009, radwan2009top, Gal_Anaby-Tavor_Trombetta_Montesi_2005}.  Consequently, these schema matching tools can effectively mitigate the risk of omitting potential results.  Some researchers have proposed algorithms that not only generate a candidate result set but also establish a search space in uncertain scenarios, as well as allocate probabilities in the context of probabilistic schema matching \cite{2008Providing, dong2009data} ,as illustrated in the top section of Table \ref{tab: Example of possible matchings result}.

As previously discussed, candidate result sets are both useful and prevalent in schema matching applications.  They enable users to achieve more accurate and comprehensive schema matching outcomes.  However, the effective utilization of candidate result sets necessitates seamless integration with databases and systems capable of accommodating probabilistic queries\cite{Detwiler_Gatterbauer_Louie_Suciu_Tarczy-Hornoch_2009, huang2009maybms}.  This requirement introduces additional complexity to the querying process and can result in increased storage costs.  Our goal is to make informed decisions at the earliest stages possible, aiming to mitigate or eliminate the propagation of uncertainty. 
 As highlighted by \cite{Popa_Velegrakis_Miller_Hernández_Fagin_2002}, human insights play a significant role in reducing uncertainty in schema matching.  A common strategy to address this uncertainty is through the involvement of human experts who provide a curated list of top-k results. 
\cite{zhang2020reducing} employs crowd-sourcing techniques to alleviate uncertainty.  \cite{Gal_2018, gal2021learning} have explored the use of learn-to-rank methods to reprioritize the candidate result sets.  Their work represents a significant advancement in the field by potentially obviating the need for human experts as the ultimate decision-makers in schema matching. However, the efficacy of these learn-to-rank approaches relies heavily on the availability of sufficient labeled data for model training—a requirement that can be difficult to fulfill in certain practical scenarios.

We observe that large language models have yet to be extensively employed in the context of uncertainty reduction for schema matching.  In \cite{pan2023rewards}, it was discovered that GPT-4, when paired with tailored prompts, achieved superior performance compared to elite crowd-sourcing workers. Consequently, this application saved a research team an estimated \$500,000 and \$20,000 hours of labor.  \cite{Gilardi_Alizadeh_Kubli_2023} demonstrate that ChatGPT outperforms crowd-workers in multiple annotation tasks, emerging as a cost-effective alternative at approximately one-fifth the expense of Amazon Mechanical Turk (MTurk), a well-known crowd-sourcing platform.  \cite{Tornberg_2023} further presents evidence that GPT-4 achieves more accurate performance than both experts and crowd workers in the textual analysis task of discerning the political leanings of Twitter users. Large Language Models (LLMs) have increasingly replaced the work of crowd-sourcing workers across various tasks.  As illustrated above, LLMs possess substantial capabilities to automate the functions traditionally performed by crowd-sourcing workers across various domains. However, there remain several challenges that need to be addressed to optimize the application of LLMs.
Our paper is dedicated to leveraging LLMs to reduce uncertainty of schema matching, a domain that remains largely unexplored.

\subsection{Challenges and Contributions}
Large Language Models require appropriately designed prompts to effectively activate their capabilities.  The design of a specific prompt tailored for the schema matching task is a crucial aspect of using LLMs.  While Large Language Models are generally more cost-effective than crowd-sourcing, the expense associated with their deployment at massive scales warrants careful consideration.  Mitigating the impact of incorrect answers from large models is also a concern.   

\begin{enumerate}
    \item \textbf{Prompt Engineering} 
    
    Proper Prompt can significantly improve the performance of LLMs in many tasks.  Relevant research on the prompt design of LLMs in schema matching is rare, especially on the correspondence verification task.
    \\

    \item \textbf{Noisy Answers} 
    
    Incorporating the confidence estimation of LLMs responses into our methodology presents a significant research challenge that requires careful consideration.
    \\
    \\
    
    \item \textbf{Budget Constraint}  

    The usage of LLMs is relatively expensive.  LLM API Service Providers set the price based on the tokens in usage.  In applications involving multiple calls to large language models, achieving the best results within a fixed budget is a key issue. 
    
\end{enumerate}

To address the aforementioned challenges, we have developed two specialized prompts for the correspondence verification task.  The Semantic-Match Prompt is designed for correspondences with clear semantics. It is characterized by its simplicity and low computational cost.  In contrast, the Abbreviation-Match Prompt is more complex and computationally expensive. However, it outperforms the Semantic-Match Prompt in handling abbreviations, which often lack explicit semantic information.  The confidence scores of verification answers have been incorporated into the updating process, thereby mitigating the impact of noisy answers.  We formulate an NP-hard problem: how to select a set of correspondences that maximizes the uncertainty reduction of candidate results within a fixed budget.  A greedy selection algorithm is established to enhance the solution efficiency.
  
The reasons for the candidate results of schema matching in practical applications.  
\begin{enumerate}
    
    \item \textbf{Complex Parameterization}  Numerous schema match algorithms require complex parameterization for optimal performance posing challenges for practitioners\cite{Koutras_Siachamis_Ionescu_Psarakis_Brons_Fragkoulis_Lofi_Bonifati_Katsifodimos_2021}.   
    
    \item \textbf{The Integration Of Results}
    Schema matching algorithms exhibit significantly different performance across various datasets and scenarios.  In \cite{Koutras_Siachamis_Ionescu_Psarakis_Brons_Fragkoulis_Lofi_Bonifati_Katsifodimos_2021}, the evaluation across fabricated dataset pairs as well as real-world data indicates that no single schema matching method consistently outperforms the rest.  For the methods of LLMs, LLMs generate different results with different prompts and hyper-parameters.  Ultimately, the integration of diverse results can enhance the overall performance of schema matching. 
    
\end{enumerate}

Our contributions are summarized as follows. 

\textbf{First} To the best of our knowledge, this is the first work to use LLMs to reduce the uncertainty of schema matching results.  Instead of using LLM to verify the result directly, our paper proposes an iterative approach based on fine-grained correspondence verification. 

\textbf{Second} To obtain more accurate verification answers, we have crafted two specific prompts for the correspondence verification task, achieving state-of-the-art results on two benchmark datasets.  To achieve the maximum reduction in uncertainty, we formulate the correspondence selection problem, which is an NP-hard problem. We further propose a greedy selection algorithm to address this challenge. 

\textbf{Third} We conducted a comprehensive experimental evaluation on the datasets, where the abbreviated attribute names were derived from real-world datasets. The results of these experiments demonstrate the effectiveness of our approach.

\section{Related Work}
\subsection{Uncertain Schema Match}
Uncertainty is common and significant in data integration systems.  As we have mentioned, 
\textbf{complex parameterization} and \textbf{the integration of results} generate multiple results.  
\cite{dong2009data, Magnani_Rizopoulos_McBrien_Montesi_2005, zhang2020reducing, Gal_2006} have researched on the uncertainty of schema match over the past 20 years.  In \cite{Gal2011}, authors thought that schema matchers are inherently uncertain and it is unrealistic to expect a single matcher to identify the correct mapping for any possible concept in a set.  In effort to increase the robustness of individual matchers in the face of matching uncertainty, researchers have turned to schema matcher ensembles\cite{2002VLDB, Gal_Sagi_2010, Magnani_Rizopoulos_Mc.Brien_Montesi_2005, Mork_Rosenthal_Korb_Samuel_2006}.  

 In \cite{dong2009data}, authors assume that a schema match system will be built on a probabilistic data model and introduce the concept of probabilistic schema mappings and analyze their formal foundations.  In \cite{zhang2020reducing}, crowd-sourcing platform is used as a basic fact provider, where workers determine whether a correspondence should exist in the schema match result or not.   In \cite{gal2021learning, Gal_2018}, learn-to-rank methods are introduced to re-rank the top-K results.   

\subsection{Data Matching and Schema Matching}
Data matching, the process of determining the equivalence of two data elements, plays a pivotal role in data integration.  Unicorn\cite{Tu_Fan_Tang_Wang_Du_Wang} align matching semantics of multiple tasks, including entity matching, entity linking, entity alignment, column type annotation, string matching, schema matching, and ontology matching.  Unicorn gets multiple SOTA results in the most of 20 datasets.  In the two datasets of schema matching, it both achieve the SOTA ,DeepMDatasets 100\% and Fabricated-datasets 89.6\% recall.

In \cite{Koutras_Siachamis_Ionescu_Psarakis_Brons_Fragkoulis_Lofi_Bonifati_Katsifodimos_2021}, schema matchers are classified into Attribute Overlap Matcher, Value Overlap Matcher, Semantic Overlap Matcher, Data Type Matcher, Distribution Matcher and Embeddings Matcher.  Different matchers have their own considerations on fuzzy matching.    Most of schema match tools use not only one Matcher.  The results of experiments also prove that there is not a single matcher which is always better than others.  Therefore, ensemble algorithm is useful in schema match area.  \cite{Zhang_Floratou_Cahoon_Krishnan_Müller_Banda_Psallidas_Patel} combines Word Embedding Featurizer, Lexical Featurizer with Bert Featurizer to generate top-K matchings suggestions for each unmapped source attribute.  Then, users would select one as the label of candidate pairs, which will be used to train the Bert model.  

There are some works about inference-only methods of LLMs on data matching task.  Agent-OM\cite{Qiang_Wang_Taylor_2024} is a novel agent-powered large language model (LLM)-based design paradigm for ontology matching(OM).  It leverages two Siamese agents for retrieval and matching, along with simple prompt-based OM tools, to address the challenges in using LLMs for OM.  ReMatch\cite{sheetrit2024rematch} is a novel schema matching method that leverages retrieval-enhanced large language models (LLMs) to address the challenges of textual and semantic heterogeneity, as well as schema size differences, without requiring predefined mappings, model training, or access to source schema data.  ReMatch represents schema elements as structured documents and retrieves semantically relevant matches, creating prompts for LLMs to generate ranked lists of potential matches. 

Although large language models have been increasingly utilized in schema matching tasks, the uncertainty of schema matching results remains a significant challenge.  Despite their potential, the application of LLMs to reduce this uncertainty has not been fully explored.  Recent studies have shown that LLMs, such as GPT-4, can outperform human crowd workers in terms of accuracy and efficiency.  However, leveraging LLMs to systematically reduce the uncertainty in schema matching results is still an underexplored area

\section{Preliminaries} 

We formally define the relevant concepts of our approach.  Let $\mathcal{CRS}$ denote a candidate result set. Table \ref{tab: Example of possible matchings result} illustrates an example of a candidate result set, displaying a single schema matching result along with its corresponding probability per row.  Let $c$ represents a correspondence. 

\hangafter 1
\hangindent 1em
\begin{definition}
\noindent \textbf{Single Candidate Result} 

Single Candidate Result 
$S_i=\{c_1, c_2,...,c_j\}$ denotes a set of correspondences that satisfy the constraint that no attribute is associated with more than one correspondence.

\end{definition}

\hangafter 1
\hangindent 1em
\begin{definition}
\noindent \textbf{Candidate Result Set}
        
Candidate Result Set is denoted by $\mathcal{CRS} = \{S_1,...,S_n\}$, a set of single candidate results.  The probability distribution of $\mathcal{CRS}$ is 
$\{ \mathbb{P}(S_1),...,\mathbb{P}(S_n) \}$, generated by its probability assignment function $\mathbb{P}$.

\end{definition}

\hangafter 1
\hangindent 1em
\begin{definition}
\label{def set and prob}
  \noindent \textbf{Correspondence Set and Probability}
The set of correspondences, denoted as $C =\{c_1,...,c_{|C|} \}$, is defined as a set of all the correspondences in $CRS$. Let $\mathbb{P}(S_i)$ and $\mathbb{P}(c_i)$ denote the probabilities of $S_i$ and $c_i$, respectively. We have
\end{definition}

\begin{equation}
  \begin{aligned}
    &C = \{c_i|S_i \in \mathcal{CRS} \ and \ c_i \in S_i \} \\
    &\mathbb{P}(c_i) = \sum_{S_i \in \mathcal{CRS}} \mathbb{P}(S_i),\ if \ c_i \in S_i  
  \end{aligned}
\end{equation}

\hangafter 1
\hangindent 1em
\begin{definition}
  \textbf{Correspondence Verification Task}
Let $c$ represent a single correspondence in $\mathcal{CRS}$. 
The correspondence verification task is to utilize the GPT-4 to verify the correctness of $c$ and response in the range of $\{True, False\}$.
\end{definition}

\hangafter 1
\hangindent 1em
\begin{definition}
  \label{def:view}
  \textbf{View}
 A \emph{View} $V$ is a truth-valued form of $\mathcal{CRS}$.  Single Candidate Result can be looked as a group of truth-valued answers or named a \emph{view} $v_i$, for all the Correspondence Verification Tasks of Correspondence Set in $\mathcal{CRS}$.  
\end{definition}

\begin{table}[!ht]
  \centering
  \caption{Example Of View}
  \label{tab:Example of Views}
  \resizebox{\columnwidth}{!}{\begin{tabular}{cccccccc}
    \hline
     & $c_1$& $c_2$ & $c_3$ & $c_4$ & $c_5$ & $c_6$ & $\mathbb{P}(s)$  \\
    \hline
    $v_1$ & True & True & True & True & False & True & 0.55 \\
    $v_2$ & True & True & False & True & True& True& 0.25 \\
    $v_3$ & False & True &True & False & False & True & 0.20 \\ 
    \hline
    \end{tabular}}
\end{table}

As illustrated in Table \ref{tab:Example of Views}, an example of \emph{View} $V$ is presented.  This \emph{View} is derived from $\mathcal{CRS}$ in Table \ref{tab: Example of possible matchings result}, in accordance with Definition \ref{def:view}.  Transforming $\mathcal{CRS}$ into \emph{View} make it easier to understand the relationship between $\mathcal{CRS}$ and Correspondence Verification Task.  For $n$ Correspondence Verification Tasks, there are $2^n$ possible truth-value answer groups or \emph{views}. For $v_i \in V$, the probability of $v_i$ being the ground truth is represented as $P(gt(V)=v_i)$.  We denote the probability $P(gt(V)=v_i)$ as $P(v_i)$ for the abbreviation.  If a correspondence $c$ is 'true' in one $v$, we say that $v$ is a positive model of $c$, which is denoted as $v \models c$.  Otherwise, if $v$ is a negative model for $c$, then $v$ does not satisfy $c$, denoted as $v \not\models c$.  Moreover, the following assertions hold for each $c \in C$, and one example is demonstrated in Table \ref{tab: Example of possible matchings result}.
\begin{equation}
  \begin{aligned}
  &P(c)=\sum_{c\in C\land v\models c}P(v) \\
  &P(\neg c)=1-\sum_{c\in C\land v\models c}P(v)=\sum_{v\in V \land v\not\models c}P(v)
  \end{aligned}
\end{equation}

Note that, the following equation is not necessarily true for $v \in V$,

\begin{equation}
  \label{eq:3}
  \begin{aligned}
  P(c)=\prod_{v_i\in\mathcal{V}}P(l_i)
  \end{aligned}
\end{equation}
where

\begin{equation}
  \begin{aligned}
  l_i=\left\{\begin{array}{ll}c_i&\text{if }v\models c_i,\\\neg c_i&\text{otherwise.}\end{array}\right.
  \end{aligned}
\end{equation}

Equation \ref{eq:3} above may not hold true, as the views in $V$ are independent. In the example of Table \ref{tab:Example of Views}, it can be observed that the probability $P(v_1)$ does not satisfy Equation \ref{eq:3}.

\hangafter 1
\hangindent 1em
\begin{definition}
  \textbf{Uncertainty Metric of Candidate Result Set}
  Given a $\mathcal{CRS}$ and its correspondence set $C$, Let the \emph{View} of $\mathcal{CRS}$ is $V$, the uncertainty of $\mathcal{CRS}$ can be represented by Shannon Entropy of $V$, denoted by $H(V)$ 
\end{definition}

\begin{equation}
  \begin{aligned}
  H(V)= -\sum_{v \in V}P(v)\log P(v),
  \end{aligned}
\end{equation}

\section{METHOD}
In this section, we present a three-stage iterative cycle algorithm to maximize the reduction of uncertainty within a fixed budget.  The three stages are correspondence selection, correspondence verification, and probabilities update.  As the figure \ref{fig: process of prompt-matcher} shows, the approach runs iteratively until the budget is out.  At each step, the correspondence selection algorithm identifies the correspondences that maximize the expected reduction in uncertainty.  The selected correspondences are filled into the prompt template as input for the LLM.  The LLM generates answers and their associated confidences.  The candidate results of schema matching update their probability distributions based on these answers and confidences.

\subsection{Correspondence Selection Problem}
To fully leverage the potential of LLMs to reduce uncertainty, we formalize the correspondence selection problem.  The proposed optimization model seeks to maximize the expected uncertainty reduction across the correspondence set subject to budget constraints, with a rigorous theoretical analysis confirming the NP-Hard nature of the correspondence selection problem.  Furthermore, we introduce a greedy selection algorithm designed to optimize computational efficiency.  We proceed to establish precise definitions and offer thorough explanations of all essential components in the subsequent discussion.

\subsubsection{Problem Formulation}
To calculate the expected uncertainty reduction of candidate results, we need to clarify the concepts of the family of answers and the probability of answers.  \textbf{Possible Answer Families} Given a correspondence set $\mathcal{T}$, we denote the set of all the possible answer families by $AS^{\mathcal{T}}$.  One possible answer is denoted by $A^{\mathcal{T}}$.

\hangafter 1
\hangindent 1em
\begin{definition} 
  \label{def: Entropy of the LLM answer families}
  \textbf{Entropy of the answer Families}
  For a given correspondence set $C$, a selected correspondence set $\mathcal{T}$, the entropy of the answer families $H(AS^{\mathcal{T}})$ is
\end{definition}

\begin{equation}
  \begin{aligned}
  H(AS^{\mathcal{T}}) = - \sum_{A^{\mathcal{T}} \in AS^{\mathcal{T}}} P(A^{\mathcal{T}}) \log P(A^{\mathcal{T}})
  \end{aligned}
\end{equation}

\hangafter 1
\hangindent 1em
\begin{definition} \textbf{Uncertainty Reduction Expectation}
  For a correspondence set $C$ and its View $V$ , a selected correspondence set $\mathcal{T}$ , the expected uncertainty of $V$ after getting answers from LLM for $\mathcal{T}$ is
\end{definition}

\begin{equation}
  \begin{aligned}
  \Delta \mathbb{H}(V) &= H(V)-\mathbb{H}(V|AS
  ^\mathcal{T}) \\
  &=H(V)+\sum_{A^{\mathcal{T}} \in AS^{\mathcal{T}}}  P(A^{\mathcal{T}})H(V|A^{\mathcal{T}})
  \end{aligned}
\end{equation}

where the $H(V|A^{\mathcal{T}})$ is the conditional uncertainty expectation with the answer $A^{\mathcal{T}}$.  To maximize the uncertainty reduction of a correspondence set, we use the uncertainty reduction expectation as our objective function.  We remove the constant term and further derive the opjective function.

\begin{table}[!ht]
  \centering
  \caption{GPT-4 Price}
  \label{tab:gpt-4 price}
  \resizebox{\columnwidth}{!}{
  \begin{tabular}{lll}
    \hline
    Model & Input & Output   \\
    \hline
    8k content& \$0.03/1k tokens & \$0.06/1k tokens \\
    32k content & \$0.06/1k tokens & \$0.12/1k tokens \\
    \hline
  \end{tabular}}
\end{table}

\begin{lemma}
\textbf{Objective Function}.
   For one View $V$, the selected correspondence set is $\mathcal{T}$.  The answer families is $AS^\mathcal{T}$.  The \textbf{Objective Function} is 
\end{lemma}

\begin{equation}
  \begin{aligned}
   -\mathbb{H}(V|AS^\mathcal{T}) &=\sum_{A^{\mathcal{T}} \in AS^{\mathcal{T}}}\sum_{v\in V}P(A^\mathcal{T})P(v|A^\mathcal{T})\log P(v|A^\mathcal{T})
  \end{aligned}
\end{equation}

\hangafter 1
\hangindent 1em
\noindent \emph{Proof.} For a selected correspondence set $\mathcal{T}$, the expected uncertainty reduction of the View is $\Delta \mathbb{H}(V|AS^\mathcal{T})$.  We have
\begin{equation}
  \begin{aligned}
    \Delta \mathbb{H}(V) &= H(V) - \mathbb{H}(V|AS^\mathcal{T})  \\
  \end{aligned}
\end{equation}
\noindent   Since $H(V)$ is constant at each round, the Objective Function can be simplified to $-\mathbb{H}(V|AS^\mathcal{T})$. 

Then, 
\begin{equation}
  \begin{aligned}
   -\mathbb{H}(V|AS^\mathcal{T}) &= \sum_{A^{\mathcal{T}} \in AS^{\mathcal{T}}} P(A^{\mathcal{T}})H(V|A^{\mathcal{T}}) \\
   &=\sum_{A^{\mathcal{T}} \in AS^{\mathcal{T}}}\sum_{v\in V}P(A^\mathcal{T})P(v|A^\mathcal{T})\log P(v|A^\mathcal{T})
  \end{aligned}
\end{equation}

Proof ends.

\noindent \textbf{Cost Metric} For all LLMs API providers, tokens are their price unit.  As Table \ref{tab:gpt-4 price} shows, OpenAI provides the prices of GPT-4 and ChatGPT on the website.  In the context of same prompt template and output format, we select the correspondence tokens as our cost metric.  Then $w$ will be denoted as the cost metric function.

\hangafter 1
\hangindent 1em
\begin{definition}
\label{def: csp}
  \textbf{Correspondence Selection Problem}. Given a View $V$, a budget $B$, and a correspondence set $C$ with its cost function $w$, our objective is to maximize $-\mathbb{H}(V|AS^\mathcal{T})$ by selecting a selected correspondence set $\mathcal{T}^*$ while ensuring that the cost of $\mathcal{T}^*$ does not exceed $B$. 
\end{definition}

\begin{equation}
  \begin{aligned}
    \mathcal{T}^*=\max_{\mathcal{T}\subseteq V, w(\mathcal{T})\leq B} -\mathbb{H}(V|AS^\mathcal{T})
  \end{aligned}
\end{equation}

After defining the Correspondence Selection Problem, we prove that this problem is NP-hard.

\begin{theorem}
  \textbf{Computation Hardness}
To solve the correspondence selection problem is an NP-hard problem.
\end{theorem}

\noindent \emph{Proof.} To establish the NP-hardness of the Correspondence Selection problem, it is sufficient to prove the NP-completeness of its decision version.  We designate the decision version as DCS (Decision Correspondence Selection). In the context of DCS, there exists a view set denoted as $V$, along with a correspondence set denoted as $C$. These sets are associated with the correspondence cost function $w()$ and the correspondence value function $value()$. Given a budget $B$ and a value $H$, the objective of DCS is to determine the existence of a subset $\mathcal{T}$ from a set $C$ that satisfies the conditions $value(\mathcal{T}) \geq H$ and $w(\mathcal{T}) \leq B$.

To establish the NP-completeness of the DCS problem, it is sufficient to prove the NP-completeness of a specific instance of DCS. Initially, the accuracy rate is set to 1.0. For a subset $\mathcal{T}$ of $C$, we let $w(\mathcal{T})=value(\mathcal{T})$, and $B=H=\frac{1}{2}\sum_{c \in C}w(c)$. If we set $X=\{w(c),\forall c \in C\}$, we can deduce that the answer to the set partitioning problem of $X$ is ``yes" if and only if the answer to the DCS is ``yes".  Since the set partitioning problem has been proven to be NP-complete by \cite{Balas_Padberg_1976}, it is reasonable to infer that the DCS problem is also NP-complete. Given that the decision version is less challenging, it can be inferred that the correspondence selection problem is an NP-hard problem.

\begin{table}[!ht]
\centering
\resizebox{\columnwidth}{!}{
\begin{tabular}{>{\raggedright\arraybackslash}p{\columnwidth}} 
\toprule
Greedy Selection with Partial Enumeration \\ 
\midrule
1: $\mathcal{T}_1 \leftarrow \emptyset, \mathcal{T}_2 \leftarrow \emptyset$ \\
2: Initialize Correspondence Set $C$, Budget $B$, \\
\quad Cost Function $w()$ \\
3: $\mathcal{T}_1 \leftarrow \arg\max_{\mathcal{T}_{1}\subseteq C, w(\mathcal{T}_1)\leq B, |\mathcal{T}_1|\leq 2} -\mathbb{H}(V|AS^{\mathcal{T}_1})$ \\
4: \textbf{for all} $\mathcal{T} \subseteq C, |\mathcal{T}|=3, w(\mathcal{T})\leq B$ \textbf{do} \\
5: \quad $C^{'} \leftarrow C \backslash \mathcal{T}$ \\
6: \quad $\mathcal{T}_{g} \leftarrow \mathcal{T}$ \\ 
7: \quad \textbf{while} $C^{'} \neq \emptyset$ \textbf{and} $w(\mathcal{T}_{g}) < B$ \textbf{do} \\
8: \quad \quad Select $c^{*} = \arg\max_{c \in C^{'}}\left\{ \frac{-\mathbb{H}(V|AS^{(\mathcal{T}_{g} \cup {c})})+\mathbb{H}(V|AS^{\mathcal{T}_g})}{w(c)} \right\}$ \\
9: \quad \quad \textbf{if} $w(\mathcal{T}_g \cup c^{*}) \le B$ \textbf{then} \\
10: \quad \quad $\mathcal{T}_g \leftarrow \mathcal{T}_g \cup \{ c^{*} \}$ \\
11: \quad \quad \textbf{end if} \\
12: \quad \quad $C^{'} \leftarrow C^{'} \backslash c^{*}$ \\
13: \quad \textbf{if} $-\mathbb{H}(V|AS^{\mathcal{T}_g}) > -\mathbb{H}(V|AS^{\mathcal{T}_2})$ \textbf{then} \\
14: \quad \quad $\mathcal{T}_2 \leftarrow \mathcal{T}_g$ \\
15: \textbf{return} $\arg\max \{w(\mathcal{T}_1), w(\mathcal{T}_2)\}$ \\
\bottomrule
\end{tabular}
}
\caption{Greedy Selection Algorithm}
\label{tab:greedy_selection}
\end{table}

\begin{table}[!ht]
    \centering
    \resizebox{\columnwidth}{!}{
    \begin{tabular}{>{\raggedright\arraybackslash}p{\columnwidth}} 
        \toprule
        Prompt-Matcher \\
        \midrule
        1: Initialize cost function $w()$, Queue $q$, and Budget \\
       \ \ \ \ $B = \text{total\_budget}/k$. \\
        2: \textbf{while} $k > 0$ \textbf{do} \\
        3: \quad Greedy Selection  $\mathcal{T}$ and push $\mathcal{T}$ into $q$. \\
        4: \quad \textbf{for} each $c$ in $q$ \textbf{do} \\
        5: \quad \quad Ask LLM and get answer $a$ and its confidence $Pr$. \\
        6: \quad \quad Update the probabilities of $V$ with $P(V|a)$. \\
        7: \quad \textbf{end for} \\
        8: \quad $k = k - 1$. \\
        9: \textbf{end while} \\
        \bottomrule
    \end{tabular}
    }
    \caption{Prompt-Matcher Algorithm}
    \label{tab:prompt_matcher}
\end{table}

\subsubsection{Greedy Selection Algorithm}
The challenge of the correspondence selection problem stems from its NP-hard nature, which necessitates an exponential complexity to obtain the optimal solution.  The time complexity of brute solution becomes impractical when dealing with large-scale data.   
Based on the \textbf{monotone submodularity of the Objective Function}, we use Algorithm 4 of \cite{horel2015notes} as base to establish an greedy algorithm with partial enumeration.  This kind of approximate algorithm was first proposed by \cite{khuller1999budgeted}. 

~\
\textbf{Partial Enumeration}  algorithm visits all the correspondence set $\mathcal{T}$, whose size is $|\mathcal{T}| \le 2$ and find the set $\mathcal{T}_1$ maximizing the objective function $\mathbb{H}(V|AS^{\mathcal{T}})$.

\textbf{Greedy Strategy} The algorithm systematically enumerates all possible subsets of the correspondence set with a cardinality of $|T|=3$, utilizing each subset as an initial starting point. Subsequently, it applies a greedy selection strategy to iteratively expand the subset until the allocated budget is fully exhausted.  The greedy strategy iteratively selects the most cost-effective correspondence at each step, as defined by the optimization criterion in line 8 of Table \ref{tab:greedy_selection}.

\subsection{Correspondence Verification}

In this section, we introduce two tailored prompt strategies: the Semantic-Match Prompt, designed for datasets with semantically meaningful attribute names, and the Abbreviation-Match Prompt, optimized for datasets featuring abbreviated attribute names.  While the Semantic-Match Prompt offers a simpler and more straightforward approach, it exhibits limitations when applied to datasets with abbreviations.  In contrast, the Abbreviation-Match Prompt demonstrates superior capability in handling such cases, albeit at the cost of increased complexity and computational expense.       

\noindent \textbf{Semantic-Match Prompt} 

\begin{figure}[!ht]
  \centering
  \includegraphics[width=\linewidth,height=\linewidth]
  {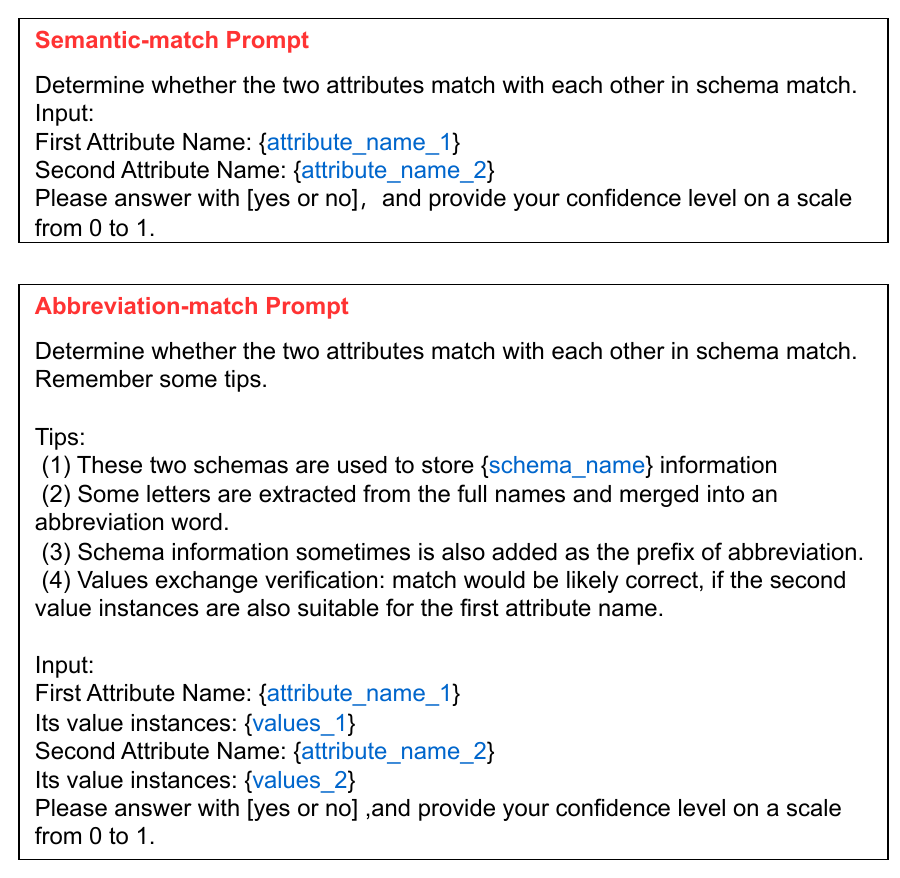}
  \captionsetup{justification=centering}
  \caption{two prompt templates, we show the prompt templates are suitable for schema matching tasks. The blue placeholders are filled with the attribute names and their values in the correspondences. schema\_name is filled with the dataset name or domain information.}
  \label{fig: allprompt}
\end{figure}

We design Semantic-Match prompt for datasets having clear semantics, as Figure \ref{fig: allprompt} shows.  The Semantic-Match Prompt is composed of three distinct parts: task description instruction,  input data, and output Indicator. By testing multiple prompt variations on the DeepMDatasets, we empirically determine that the Semantic-Match Prompt is highly effective for datasets with characteristics similar to DeepMDatasets. 
\begin{enumerate}
    \item \textbf{task description instruction}, is "Determine whether the two attributes match with each other in schema match".  The goal is to describe the task instructions clearly.
    \item \textbf{input data}, the attributes of DeepMDatasets own clear semantics.  Therefore, we take the names of target schema attributes and sources schema attributes as input.  
    \item \textbf{output indicator},  The goal is to instruct the output format of LLM.
\end{enumerate}
The structure of Semantic-Match Prompt is simple and cost-effective.  When the attribute names have clear semantics in schema matching datasets, GPT-4 with Semantic-Match Prompt template has good performances.

\ \\
\noindent \textbf{Abbreviation-Match Prompt}

In real-world datasets, abbreviation is the core challenge.  The Semantic-Match Prompt exhibits limited effectiveness in verifying abbreviated correspondences.  To address this limitation, we designed the Abbreviation-Match Prompt, which is grounded in the core principles of classic schema matching algorithms.  As illustrated in Figure \ref{fig: allprompt}, Tip (1) incorporates domain-specific information into the LLM to enhance context awareness.  Tips (2) and (3) are inspired by the COMA algorithm \cite{Do_Rahm_2002}, guiding the LLM to recognize and apply rules for abbreviation generation.  Furthermore, Tip (4) draws on the design philosophy of the "Value Overlap Matcher," which evaluates attribute similarity based on value overlap.  Specifically, we instruct the LLM to exchange attribute values and assess their similarity, thereby improving matching accuracy.

As the figure \ref{fig: allprompt} shows, the Abbreviation-Match Prompt template has five blue placeholders, \textbf{schema\_name} is filled with the domain name of datasets or the name schema matching.  The subset attribute names of correspondence are filled into the \textbf{attribute\_name} placeholders seperately.  And, we directly fill the \textbf{values} placeholder with the first three values of the attribute.

\subsection{Probability Update}
In this section, we formulate the method to update the probabilities of candidate results to reduce uncertainty.  We take the candidate result set in Table \ref{tab: Example of possible matchings result} as an example.  Initially, the correspondence set $c_2=[Email, Email Address]$ is selected.  We make an assumption that the statement "I:in EmployeeInfo and Employee $c_2$=[\textit{Email}, \textit{Email Address}] is right correspondence" is a response from LLM with a confidence of 80\%.  Then, the confidence of the View will be updated using the following equation:

\begin{equation}
    \begin{aligned}
    &Pr(v_1|e:I \ from \ LLM) = Pr(v_1)Pr(e|v_1)/Pr(e) \\
    &=\frac{Pr(v_1)Pr(LLM~is~correct)}{Pr(I)Pr(C~is~correct)+(\neg Pr(I))Pr(C~is~incorrect)} \\
    &=\frac{0.55*0.80}{0.80*0.80+0.20*0.20} = 0.647
    \end{aligned}
\end{equation}

Similarly, $Pr(v_2|e)=0.294$ and $Pr(v_3|e)=0.059$.  The uncertainty $H(V)$ has been reduced from 0.997 to 0.351.  In the subsequent parts, we

\textbf{Computation of answer set probability}
Given a correspondence set $C$ , a selected correspondence set $\mathcal{T} \subset C$, the confidences of the $A^{\mathcal{T}} $is $\{Pr_1, Pr_2,...,Pr_{|A^{\mathcal{T}}|} \}$ and a view is $v$,
the conditional probability of receiving $A^{\mathcal{T}}$  given a view $v$ is

\begin{equation}
  \label{eq: p_a_t_v}
  \begin{aligned}
  & P(A^{\mathcal{T}}|v)=\{P(a_1|v),P(a_2|v),...,P(a_{|A^{\mathcal{T}}|} |v) \} \\
  \end{aligned}
\end{equation}
where,
\begin{equation}
  \begin{aligned}
   P(a_i|v)=\left\{\begin{array}{ll}Pr_i*P(v)+(1-Pr_i)*(1-P(v))&\text{if }v\models a_i\\
   Pr_i*(1-P(v))+(1-Pr_i)*P(v) &\text{if }v\not\models a_i\end{array}\right.
  \end{aligned}
\end{equation}

and the probability of $A^{\mathcal{T}}$ is
\begin{equation}
  \label{eq: p_a_t}
  \begin{aligned}
    &P(A^{\mathcal{T}})=\{P(a_1), P(a_2),...,P(a_i) \} \\
  \end{aligned}
\end{equation}

where,
\begin{equation}
  \label{eq: p_a_t_eq}
  \begin{aligned}
    &P(a_i)=P(a_i|v)*P(v) \\
  \end{aligned}
\end{equation}

\noindent \textbf{Updating Probability Distribution} Supposing that GPT-4 has confidence $Pr$ of the correspondence verification task in $V$.  For a selected correspondence set $\mathcal{T}$, suppose the answer set is $A^{\mathcal{T}}$, the probability of view $v \in V$ will be updated to

\begin{equation}
  \label{eq: p(v|a_t)}
  \begin{aligned}
    &P(v|A^{\mathcal{T}})=\{P(v|{a_1}), P(v|{a_2}),..,P(v|{a_i}),..,P(v|{a_{|A^\mathcal{T}|}})  \}\\
  \end{aligned}
\end{equation}

Where,

\begin{equation}
  \label{eq: p_v_a_t}
  \begin{aligned}
    &P(v|a_i)=\frac{P(v)\cdot P(a_i|v)}{P(a_i)} \\
  \end{aligned}
\end{equation}

\subsection{Prompt-Matcher}
Prompt-Matcher is a framework that helps people to reduce the uncertainty of candidate result set and rank the best result to the front too.
Prompt-Matcher is established on the loop of correspondence selection, the verification of LLM with prompt and the probability distribution updating of View.  At each iteration, prompt-matcher get the more reliable verified answers from LLM to reassign the probabilities to various schema match results.

\textbf{Initialization} $B$ is initialized by $total\_budget / k$, which is used as the budget of each round.  $k$ is the number of rounds in the loop, which is a parameter decided by users.  To test the performance of prompt-matcher, we choose one budget larger than the cost of all correspondences and larger than three times of the mean cost.  In fact, the budget can be set at any value.  Due to greedy algorithm with partial enumeration will brutely search when the number of the selected correspondence set is less than 3, we recommend to avoid setting a large budget at each round.  The large budget at one round will result in a quite high time cost.   

\textbf{Note that} greedy algorithm with partial enumeration can provide a $(1-1/e)$ approximate solution at each round.  Then, the solution of $k$ rounds will also be $(1-1/e)$.  In fact, if there are some remaining budget at each round, it will passed to next round.  So the solution of our greedy selection will have higher ratio than $(1-1/e)$.

\begin{figure}[ht!]
  \centering
  \includegraphics[width=\linewidth,height=0.75\linewidth]
  {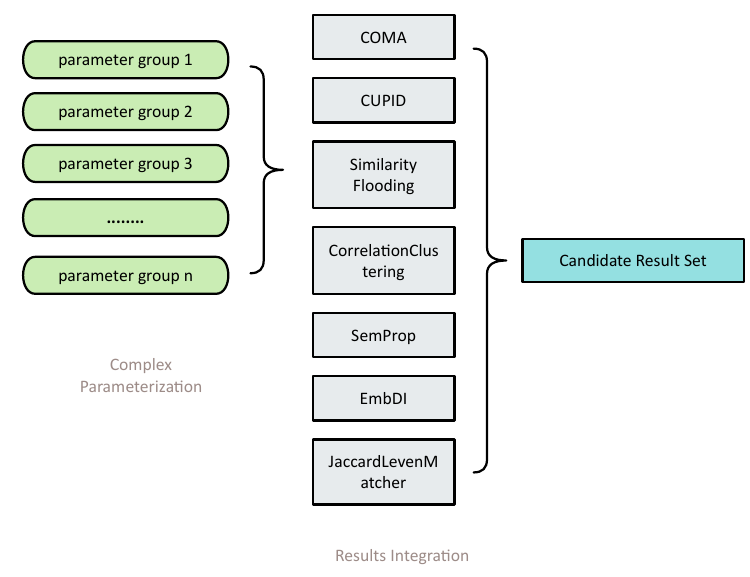}
  \captionsetup{justification=centering}
  \caption{Candidate Result Set, We systematically demonstrate the procedural workflow for candidate result set generation.}
  \label{fig: crs}
\end{figure}

\begin{figure*}[ht!]
  \centering
  \includegraphics[width=0.95\linewidth,height=0.2\linewidth]
  {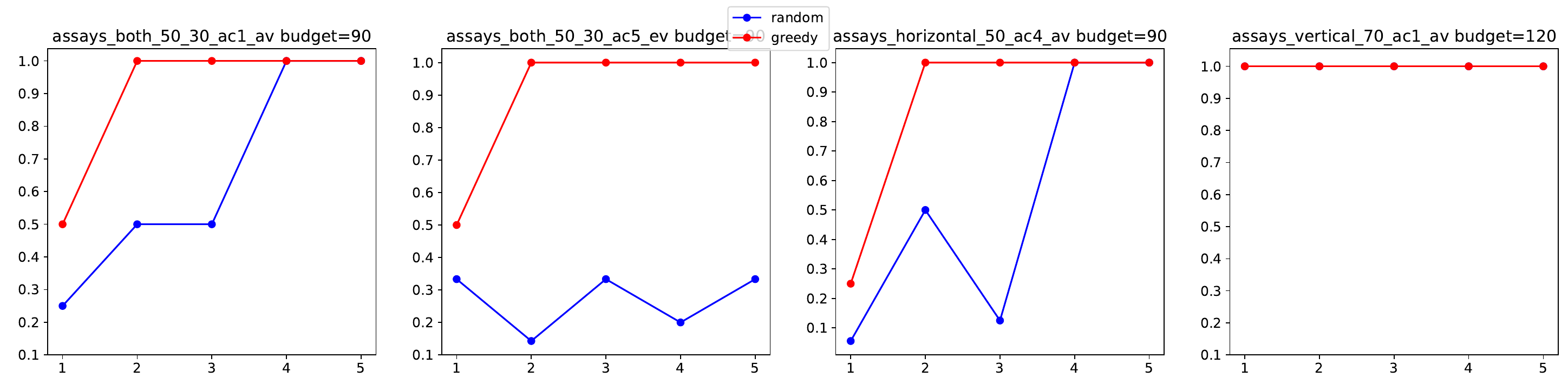}
  \captionsetup{justification=centering}
  \caption{Assays Experiment Result: MRR}
  \label{fig: assays experiment result}
\end{figure*}
\begin{figure*}[ht!]
  \centering
  \includegraphics[width=0.95\linewidth,height=0.2\linewidth]
  {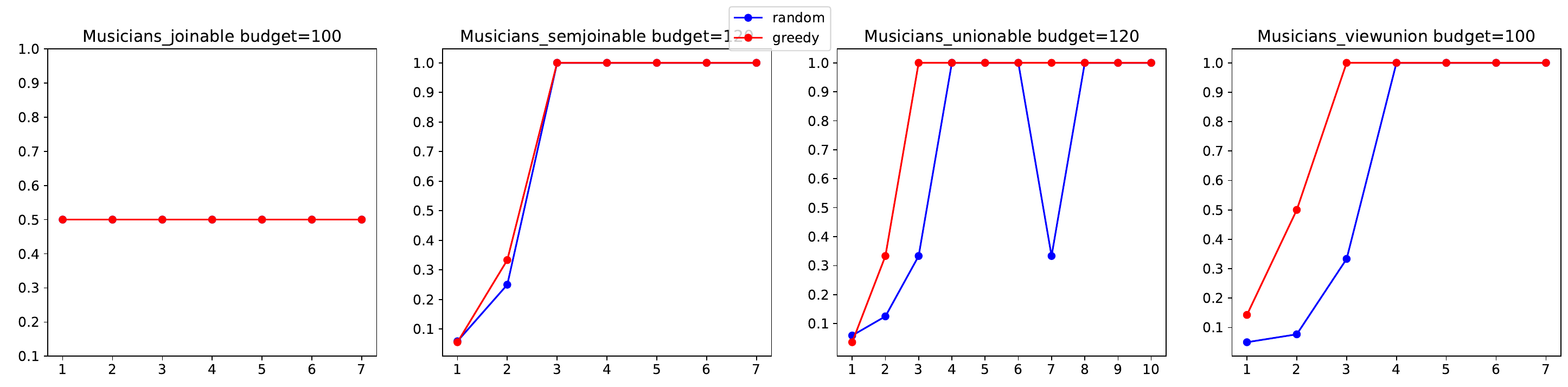}
  \captionsetup{justification=centering}
  \caption{Musician Experiment Result: MRR}
  \label{fig: musician experiment result}
\end{figure*}
\begin{figure*}[ht!]
  \centering
  \includegraphics[width=0.95\linewidth,height=0.2\linewidth]
  {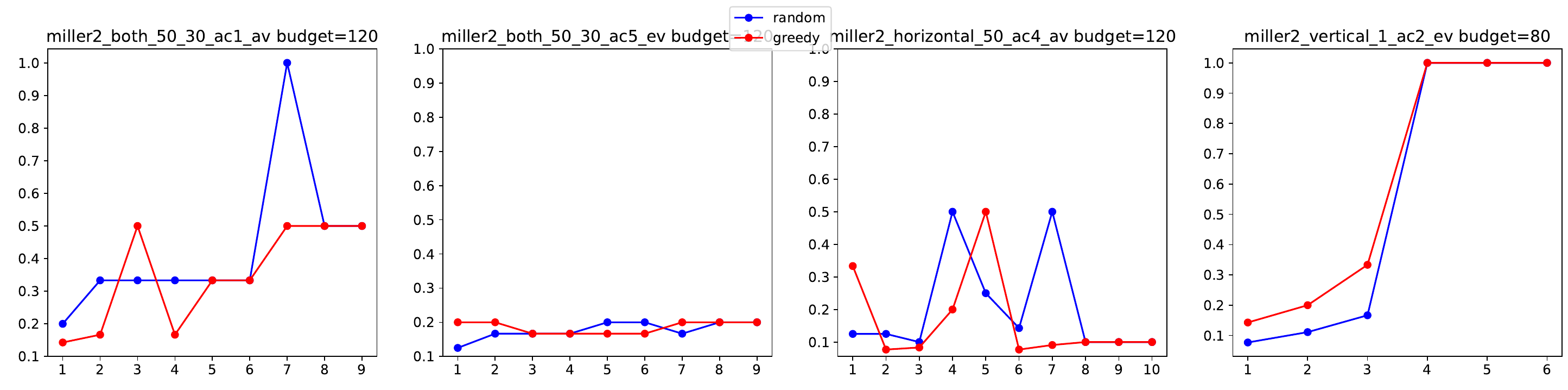}
  \captionsetup{justification=centering}
  \caption{Miller2 Experiment Result: MRR}
  \label{fig: miller2 experiment result}
\end{figure*}
\begin{figure*}[ht!]
  \centering
  \includegraphics[width=0.95\linewidth,height=0.2\linewidth]
  {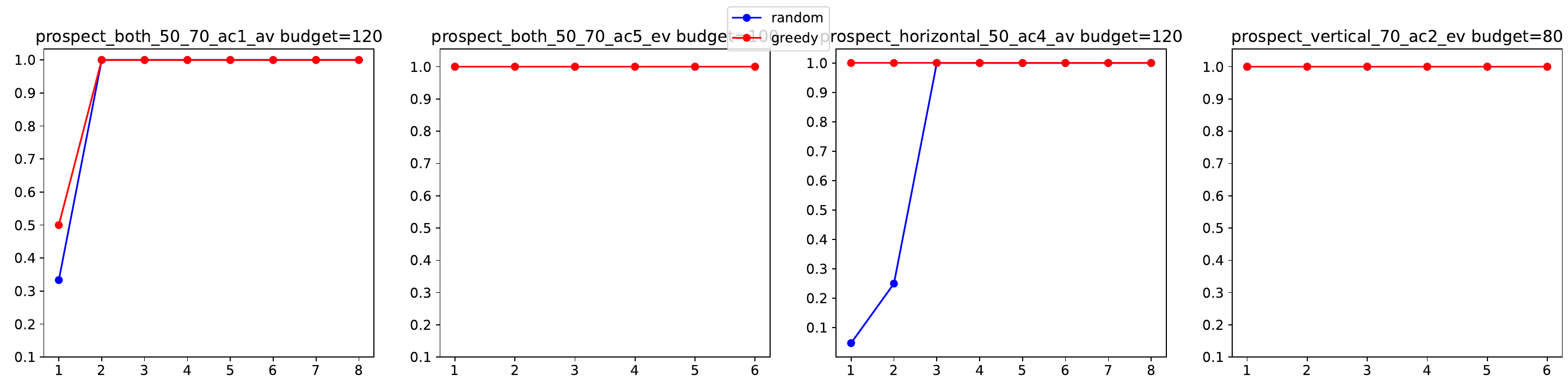}
  \captionsetup{justification=centering}
  \caption{Prospect Experiment Result: MRR}
  \label{fig: prospect experiment result}
\end{figure*}

\begin{figure*}[ht!]
  \centering
  \includegraphics[width=0.95\linewidth,height=0.25\linewidth]
  {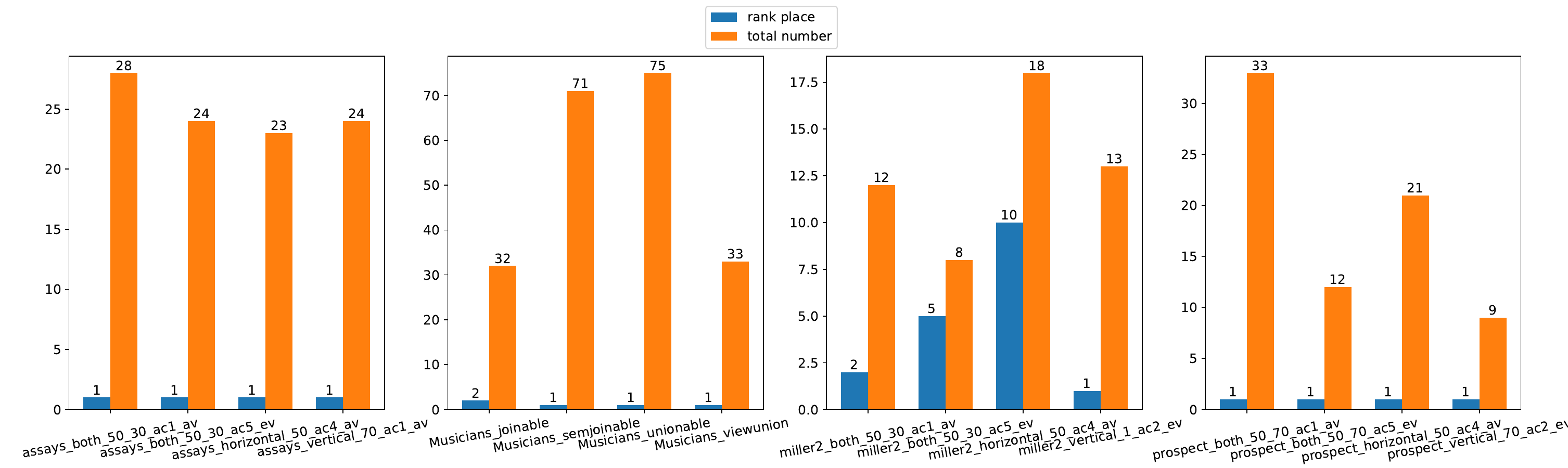}
  \captionsetup{justification=centering}
  \caption{Rank Place vs Total number}
  \label{fig: comparative result}
\end{figure*}

\begin{figure*}[ht!]
  \centering
  \includegraphics[width=0.95\linewidth,height=0.25\linewidth]
  {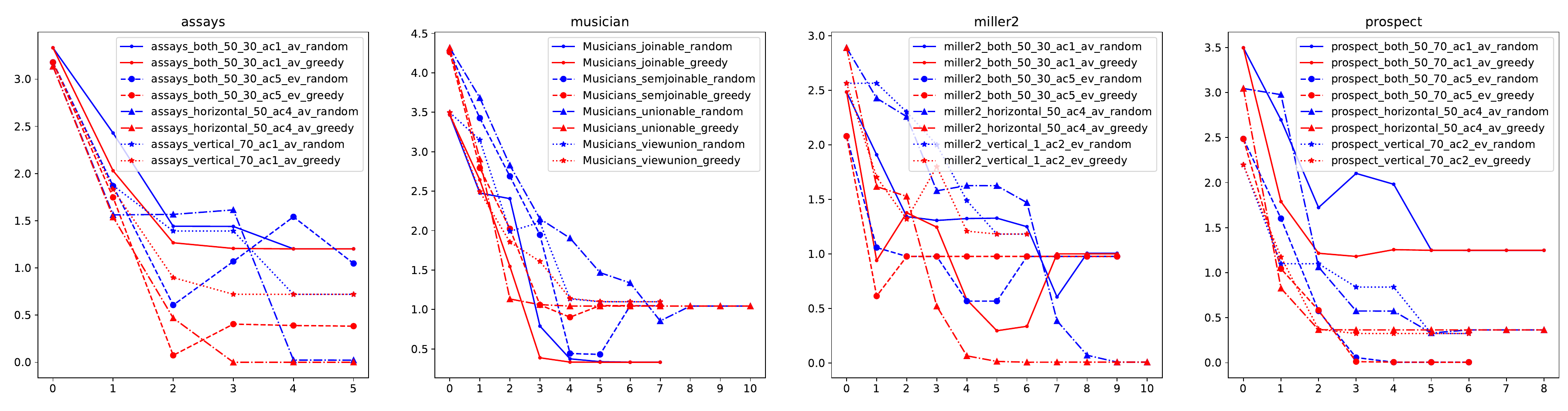}
  \captionsetup{justification=centering}
  \caption{uncertainty reduction}
  \label{fig: uncertainty reduction}
\end{figure*}

\section{Experiment Results}
\subsection{Datasets Information}
Fabricated-Dataset
\cite{Koutras_Siachamis_Ionescu_Psarakis_Brons_Fragkoulis_Lofi_Bonifati_Katsifodimos_2021} is composed of many real-world datasets across multiple domains, having abbreviation attribute names.  This dataset is designed to evaluate the performance and effectiveness of data discovery algorithms and schema matching algorithms. 

The DeepMDataset\cite{Mudgal_Li_Rekatsinas_Doan_Park_Krishnan_Deep_Arcaute_Raghavendra_2018}  includes the following seven real-world datasets from diverse industries:
Amazon-Google: A software product dataset with 11,460 labeled tuple pairs.
BeerAdvocate-RateBeer: A beer dataset with 450 labeled tuple pairs.
DBLP-ACM: A citation dataset with 12,363 labeled tuple pairs.
DBLP-Scholar: A citation dataset with 28,707 labeled tuple pairs.
Fodors-Zagats: A restaurant dataset with 946 labeled tuple pairs.
iTunes-Amazon: A music dataset with 539 labeled tuple pairs.
Walmart-Amazon: An electronics dataset with 10,242 labeled tuple pairs.
These datasets cover a wide range of domains, including e-commerce, academic citations, music, and restaurants, and are widely used to evaluate schema matching algorithms.  
\noindent\textbf{Note that} We conducted our experiments on the test dataset of above datasets.

\subsection{The Experiment of Correspondence Verification}
As the table \ref{tab: comprehensive evaluations} shows, we conduct compratative experiments on the models which are used in the classification task.  With Semantic-Match Prompt, GPT-4 get 100\% accuracy on the test dataset of DeepMDataset, which is equal to the SOTA performance.  However, it gets just 73\% recall rate on Fabricated-Datasets.  It shows that the limit of Semantic-Match Prompt on abbreviation attribute names.  The Abbreviation-Match Prompt with GPT-4 get 91.8\% accuracy in Fabricated-Dataset, which is a new SOTA result.  Abbreviation-Match Prompt also get 99.35\% accuracy on DeepMDataset.  Although Abbreviation-Match Prompt is more complex and expensive than Semantic-Match Prompt, it has more robust performances in our experiments.  

We also carry out ablation experiments on Abbreviation-Match Prompt to explore the effect of each tip.  As table \ref{tab: abbreviation-match} shows, tips (1)(2)(3) passing the regulation of bring large increase at accuracy.  Each tip has good effect on the correspondence verification.
 \begin{table}
  \centering
  \caption{Fabricated-Dataset Prompt Ablation Experiments}
  \label{tab: abbreviation-match}
  \resizebox{\columnwidth}{!}{
  \begin{tabular}{lcc}
    \hline
    Prompt &  Accuracy \\
    \hline
     Semantic-match Prompt& 70\% \\
    Semantic-match Prompt+(4) tips& 79.3\%\\
    Semantic-match Prompt+(1)(2)(3) tip& 90.9\% \\
    Semantic-match Prompt+(1)(2)(3)(4) tips& 91.8\% \\
    \hline
  \end{tabular}}
\end{table}

 \begin{table}
  \centering
  \caption{The Results (Accuracy) on Fabricated-Datasets and DeepMDatasets}
  \label{tab: comprehensive evaluations}
  \resizebox{\columnwidth}{!}{
  \begin{tabular}{|c|c|c|}
    \hline
    &DeepMDatasets &Fabricated-Dataset \\
    \hline
    BERT & 88.89 & 77.85  \\
    RoBERTa & 88.89 & 77.72 \\
    DistilBERT & 92.59 & 42.96 \\
    DistilRoBERTa & 92.59& 77.48 \\
    XLNet & \textbf{100} & 70.99 \\
    MPNet & 88.89 & 70.62 \\
    DeBERTa & 96.3 & 89.6 \\
    Unicorn & 96.3 & 89.6 \\ 
    Unicorn++ & 96.3 & 89.35 \\
    Previous-SOTA &100 & 89.6 \\
    \hline
    Semantic-match &\textbf{100} &70.0 \\
    Abbreviation-match & 99.35 & \textbf{91.8} \\
    \hline 
  \end{tabular}}
\end{table}

\subsection{The Experiments On Prompt-Matcher}

\subsubsection{Settings}
To more comprehensively test our algorithm, we choose 16 datasets from Fabricated-Dataset, which are fabricated from 4 real-world datasets (musician, assays, prospect, miller2).  As Figure \ref{fig: crs} shows, we collect candidate result set from 7 schema matching algorithms with multiple groups of parameters.  We do contrast experiments between random algorithm and greedy selection.  Our codes and the details of experiment can be found at \cite{promptMatcher}.  we take \textbf{MRR}(Mean Reciprocal Rank) as the evaluation metric.  \textbf{MRR} can represent the location of optimal result.  The equation of computation is $MRR=\frac{1}{rank}$, where $rank$ is the location of the optimal result.  We take the F1-score of the correspondences in one candidate result as its metric.  

\textbf{The effect of Prompt-Matcher}
In figure \ref{fig: assays experiment result}, \ref{fig: musician experiment result}, \ref{fig: miller2 experiment result}, \ref{fig: prospect experiment result}, we can find that Prompt-Matcher with random selection or greedy selection can rank the best schema match result to the first place in the 12/16 (\textbf{75\%}) experiments.  Further, 14/16 (\textbf{87.5\%}) experiments can rank the best result to the second place.  The final ranking place is highly related to the accuracy of GPT-4.  Although we achieve the SOTA result, 91.8\% accuracy and recall in the fabricated datasets, it can not ensure Prompt-Matcher can rank the best schema match result to the first place.  As figure \ref{fig: comparative result} shows, the experiment of miller2\_both\_50\_30\_ac5\_ev dataset has a bad result.  We check the accuracy of GPT-4 on this dataset.  The result is 21/37 (56.7\%), which is far less than 91.8\%.  GPT-4 with Abbreviation-match prompt has poor performance on this dataset.  So it is quite necessary to evaluate the accuracy of GPT-4 on the dataset.  If the evaluation result of GPT-4 on your dataset is quite good, top-3 or 5 can cover the best schema match result.  Another bad case is the miller2\_horizontal\_50\_ac4\_av.  We find that the accuracy of GPT-4 on this dataset is quite high, 54/55 (98.1\%).  However,the f1 scores of the candidate results in $\mathcal{CRS}$ are really low.  The highest one is 0.169, which is a bad performance.  When the results in $\mathcal{CRS}$ are all far from the ground truth, prompt-matcher can not get a good result.     

\textbf{Greedy selection is more effective than random selection within the same budget} As depicted in Figures \ref{fig: assays experiment result}, \ref{fig: musician experiment result}, \ref{fig: miller2 experiment result}, and \ref{fig: prospect experiment result}, the MRR metric for random selection exhibits a slower rate of increase compared to our greedy selection algorithm as the budget expands.  The results show that our greedy selection can work better than random selection when the budget is not enough.  With the exception of the first three line charts in Figure \ref{fig: miller2 experiment result}, the performance of the greedy selection approach is at least as good as that of random selection. Moreover, the optimal outcome of the greedy selection experiment is closer to the top-performing result at equivalent budget costs.  Prompt-Matcher, integrated with a greedy selection strategy for the Correspondence Selection Problem (CSP), proves to be effective. It expedites the ranking of the optimal schema match result by reducing uncertainty within the Candidate Result Set. As illustrated in Figure \ref{fig: uncertainty reduction}, greedy selection facilitates a more rapid reduction in uncertainty compared to random selection across the majority of the 16 experiments.  This phenomenon is consistent with the phenomenon in figure \ref{fig: assays experiment result}, \ref{fig: miller2 experiment result}, \ref{fig: prospect experiment result}, \ref{fig: musician experiment result}.  Prompt-Matcher with greedy selection can make a faster uncertainty reduction of $\mathcal{CRS}$, and rank the best schema match result to the front faster.  The brute algorithm is capable of obtaining an optimal solution to the Correspondence Selection Problem (CSP). However, when compared to the brute force algorithm, the advantages of greedy selection merit consideration, particularly in terms of efficiency.

\textbf{Greedy selection need less time cost than brute algorithm} We conducted a comparative analysis of the time complexity between our "greedy selection with partial enumeration" algorithm and the brute force approach. As depicted in Table \ref{tab: time cost experiment}, the brute force algorithm is significantly more time-consuming than our proposed method. Moreover, the time complexity of the brute force algorithm exhibits a rapid increase as the budget expands, as illustrated in the table. This is attributable to its exponential time complexity. Consequently, as the scale increases, the time cost of the brute force algorithm escalates, surpassing that of greedy selection with partial enumeration.  

In most cases, prompt-matcher is quite effective to rank the best schema match result to the front of candidate result set.  When the budget can not cover all correspondences, Greedy selection can result in a better uncertainty reduction and ranking effect than random selection.  When time cost is not concerned and budget is insufficient, brute algorithm is suitable.   

\begin{table}
  \centering
  \large
  \begin{tabular}{|c|c|c|c|c|}
    \hline
    Budget & Brute Algorithm& Greedy Selection  \\
    \hline
    120&  6.81s & 2.2s  \\
    \hline
     150& 43.11s & 5.26s \\
    \hline
    180 & 188.39s & 12.16 \\
    \hline
    210 & 690.86s & 30.11s \\
    \hline
  \end{tabular}
  \caption{Time Cost of Musician\_unionable dataset, The computational advantage of greedy selection over brute-force algorithms becomes increasingly pronounced with expanding budget sizes. (The computational cost depends exclusively on the budget scale, while exhibiting complete independence from the data domain.).}
  \label{tab: time cost experiment}
\end{table}

\section{Conclusion and future work}
In this paper, we introduce Prompt-Matcher, a novel approach that leverages Large Language Models to reduce uncertainty in candidate result sets. Under a fixed budget constraint, we formulate the Correspondence Selection Problem, aiming to maximize uncertainty reduction within the given budget. We demonstrate that the Correspondence Selection Problem is NP-hard and propose an efficient approximation algorithm with a guaranteed approximation ratio of 
$(1-1/e)$.

Uncertainty is a pervasive challenge in various components of modern data integration systems, such as entity resolution, schema matching, truth discovery, and name disambiguation. We argue that integrating LLMs into these systems can significantly mitigate uncertainty and enhance decision-making processes. Looking ahead, we plan to further optimize the efficiency of our correspondence selection algorithm and incorporate self-consistency inference mechanisms from LLMs to improve the accuracy of correspondence verification.




\bibliographystyle{cas-model2-names}
\bibliography{main}

\begin{thebibliography}{36}
\expandafter\ifx\csname natexlab\endcsname\relax\def\natexlab#1{#1}\fi
\providecommand{\url}[1]{\texttt{#1}}
\providecommand{\href}[2]{#2}
\providecommand{\path}[1]{#1}
\providecommand{\DOIprefix}{doi:}
\providecommand{\ArXivprefix}{arXiv:}
\providecommand{\URLprefix}{URL: }
\providecommand{\Pubmedprefix}{pmid:}
\providecommand{\doi}[1]{\href{http://dx.doi.org/#1}{\path{#1}}}
\providecommand{\Pubmed}[1]{\href{pmid:#1}{\path{#1}}}
\providecommand{\bibinfo}[2]{#2}
\ifx\xfnm\relax \def\xfnm[#1]{\unskip,\space#1}\fi
\bibitem[{Balas and Padberg(1976)}]{Balas_Padberg_1976}
\bibinfo{author}{Balas, E.}, \bibinfo{author}{Padberg, M.W.}, \bibinfo{year}{1976}.
\newblock \bibinfo{title}{Set partitioning: A survey}.
\newblock \bibinfo{journal}{SIAM Review} \bibinfo{volume}{18}, \bibinfo{pages}{710–760}.
\newblock \URLprefix \url{http://dx.doi.org/10.1137/1018115}, \DOIprefix\doi{10.1137/1018115}.
\bibitem[{Bernstein et~al.(2011)Bernstein, Madhavan and Rahm}]{bernstein2011generic}
\bibinfo{author}{Bernstein, P.A.}, \bibinfo{author}{Madhavan, J.}, \bibinfo{author}{Rahm, E.}, \bibinfo{year}{2011}.
\newblock \bibinfo{title}{Generic schema matching, ten years later}.
\newblock \bibinfo{journal}{Proceedings of the VLDB Endowment} \bibinfo{volume}{4}, \bibinfo{pages}{695--701}.
\bibitem[{Bonifati and Velegrakis(2011)}]{Bonifati_Velegrakis_2011}
\bibinfo{author}{Bonifati, A.}, \bibinfo{author}{Velegrakis, Y.}, \bibinfo{year}{2011}.
\newblock \bibinfo{title}{Schema matching and mapping: from usage to evaluation}, in: \bibinfo{booktitle}{Proceedings of the 14th International Conference on Extending Database Technology}.
\newblock \DOIprefix\doi{10.1145/1951365.1951431}.
\bibitem[{Detwiler et~al.(2009)Detwiler, Gatterbauer, Louie, Suciu and Tarczy-Hornoch}]{Detwiler_Gatterbauer_Louie_Suciu_Tarczy-Hornoch_2009}
\bibinfo{author}{Detwiler, L.}, \bibinfo{author}{Gatterbauer, W.}, \bibinfo{author}{Louie, B.}, \bibinfo{author}{Suciu, D.}, \bibinfo{author}{Tarczy-Hornoch, P.}, \bibinfo{year}{2009}.
\newblock \bibinfo{title}{Integrating and ranking uncertain scientific data}, in: \bibinfo{booktitle}{2009 IEEE 25th International Conference on Data Engineering}.
\newblock \URLprefix \url{http://dx.doi.org/10.1109/icde.2009.209}, \DOIprefix\doi{10.1109/icde.2009.209}.
\bibitem[{Do(2002)}]{2002VLDB}
\bibinfo{author}{Do, H.H.}, \bibinfo{year}{2002}.
\newblock \bibinfo{title}{Vldb '02: Proceedings of the 28th international conference on very large databases || coma — a system for flexible combination of schema matching approaches} , \bibinfo{pages}{610--621}.
\bibitem[{Do and Rahm(2002)}]{Do_Rahm_2002}
\bibinfo{author}{Do, H.H.}, \bibinfo{author}{Rahm, E.}, \bibinfo{year}{2002}.
\newblock \bibinfo{title}{COMA: a system for flexible combination of schema matching approaches}.
\newblock p. \bibinfo{pages}{610–621}.
\newblock \URLprefix \url{http://dx.doi.org/10.1016/b978-155860869-6/50060-3}, \DOIprefix\doi{10.1016/b978-155860869-6/50060-3}.
\bibitem[{Dong et~al.(2009a)Dong, Halevy and Yu}]{Dong_Halevy_Yu_2009}
\bibinfo{author}{Dong, X.L.}, \bibinfo{author}{Halevy, A.}, \bibinfo{author}{Yu, C.}, \bibinfo{year}{2009}a.
\newblock \bibinfo{title}{Data integration with uncertainty}.
\newblock \bibinfo{journal}{The VLDB Journal} \bibinfo{volume}{18}, \bibinfo{pages}{469–500}.
\newblock \URLprefix \url{http://dx.doi.org/10.1007/s00778-008-0119-9}, \DOIprefix\doi{10.1007/s00778-008-0119-9}.
\bibitem[{Dong et~al.(2009b)Dong, Halevy and Yu}]{dong2009data}
\bibinfo{author}{Dong, X.L.}, \bibinfo{author}{Halevy, A.}, \bibinfo{author}{Yu, C.}, \bibinfo{year}{2009}b.
\newblock \bibinfo{title}{Data integration with uncertainty}.
\newblock \bibinfo{journal}{The VLDB Journal} \bibinfo{volume}{18}, \bibinfo{pages}{469--500}.
\bibitem[{flyingwaters(2023)}]{promptMatcher}
\bibinfo{author}{flyingwaters}, \bibinfo{year}{2023}.
\newblock \bibinfo{title}{prompt\_matcher}.
\bibitem[{Gal(2006)}]{Gal_2006}
\bibinfo{author}{Gal, A.}, \bibinfo{year}{2006}.
\newblock \bibinfo{title}{Managing uncertainty in schema matching with top-k schema mappings}.
\newblock p. \bibinfo{pages}{90–114}.
\newblock \DOIprefix\doi{10.1007/118030345}.
\bibitem[{Gal(2011)}]{Gal2011}
\bibinfo{author}{Gal, A.}, \bibinfo{year}{2011}.
\newblock \bibinfo{title}{Uncertain schema matching: the power of not knowing}, in: \bibinfo{booktitle}{Proceedings of the 20th ACM international conference on Information and knowledge management}.
\newblock \DOIprefix\doi{10.1145/2063576.2064039}.
\bibitem[{Gal et~al.(2005)Gal, Anaby-Tavor, Trombetta and Montesi}]{Gal_Anaby-Tavor_Trombetta_Montesi_2005}
\bibinfo{author}{Gal, A.}, \bibinfo{author}{Anaby-Tavor, A.}, \bibinfo{author}{Trombetta, A.}, \bibinfo{author}{Montesi, D.}, \bibinfo{year}{2005}.
\newblock \bibinfo{title}{A framework for modeling and evaluating automatic semantic reconciliation}.
\newblock \bibinfo{journal}{The VLDB Journal} \bibinfo{volume}{14}, \bibinfo{pages}{50–67}.
\newblock \URLprefix \url{http://dx.doi.org/10.1007/s00778-003-0115-z}, \DOIprefix\doi{10.1007/s00778-003-0115-z}.
\bibitem[{Gal et~al.(2018)Gal, Roitman and Shraga}]{Gal_2018}
\bibinfo{author}{Gal, A.}, \bibinfo{author}{Roitman, H.}, \bibinfo{author}{Shraga, R.}, \bibinfo{year}{2018}.
\newblock \bibinfo{title}{Heterogeneous data integration by learning to rerank schema matches}.
\newblock \bibinfo{journal}{Industrial Conference on Data Mining} \DOIprefix\doi{10.1109/icdm.2018.00118}.
\bibitem[{Gal et~al.(2021)Gal, Roitman and Shraga}]{gal2021learning}
\bibinfo{author}{Gal, A.}, \bibinfo{author}{Roitman, H.}, \bibinfo{author}{Shraga, R.}, \bibinfo{year}{2021}.
\newblock \bibinfo{title}{Learning to rerank schema matches}.
\newblock \bibinfo{journal}{IEEE Transactions on Knowledge \& Data Engineering} \bibinfo{volume}{33}, \bibinfo{pages}{3104--3116}.
\bibitem[{Gal and Sagi(2010)}]{Gal_Sagi_2010}
\bibinfo{author}{Gal, A.}, \bibinfo{author}{Sagi, T.}, \bibinfo{year}{2010}.
\newblock \bibinfo{title}{Tuning the ensemble selection process of schema matchers}.
\newblock \bibinfo{journal}{Information Systems} \bibinfo{volume}{35}, \bibinfo{pages}{845–859}.
\newblock \DOIprefix\doi{10.1016/j.is.2010.04.003}.
\bibitem[{Gilardi et~al.(2023)Gilardi, Alizadeh and Kubli}]{Gilardi_Alizadeh_Kubli_2023}
\bibinfo{author}{Gilardi, F.}, \bibinfo{author}{Alizadeh, M.}, \bibinfo{author}{Kubli, M.}, \bibinfo{year}{2023}.
\newblock \bibinfo{title}{Chatgpt outperforms crowd-workers for text-annotation tasks} .
\bibitem[{Horel(2015)}]{horel2015notes}
\bibinfo{author}{Horel, T.}, \bibinfo{year}{2015}.
\newblock \bibinfo{title}{Notes on greedy algorithms for submodular maximization}.
\newblock \bibinfo{journal}{Lecture Notes, Available at https://thibaut. horel. org/submodularity/notes/02-12. pdf} .
\bibitem[{Huang et~al.(2009)Huang, Antova, Koch and Olteanu}]{huang2009maybms}
\bibinfo{author}{Huang, J.}, \bibinfo{author}{Antova, L.}, \bibinfo{author}{Koch, C.}, \bibinfo{author}{Olteanu, D.}, \bibinfo{year}{2009}.
\newblock \bibinfo{title}{Maybms: a probabilistic database management system}, in: \bibinfo{booktitle}{Proceedings of the 2009 ACM SIGMOD International Conference on Management of data}, pp. \bibinfo{pages}{1071--1074}.
\bibitem[{Khuller et~al.(1999)Khuller, Moss and Naor}]{khuller1999budgeted}
\bibinfo{author}{Khuller, S.}, \bibinfo{author}{Moss, A.}, \bibinfo{author}{Naor, J.S.}, \bibinfo{year}{1999}.
\newblock \bibinfo{title}{The budgeted maximum coverage problem}.
\newblock \bibinfo{journal}{Information processing letters} \bibinfo{volume}{70}, \bibinfo{pages}{39--45}.
\bibitem[{Koutras et~al.(2021)Koutras, Siachamis, Ionescu, Psarakis, Brons, Fragkoulis, Lofi, Bonifati and Katsifodimos}]{Koutras_Siachamis_Ionescu_Psarakis_Brons_Fragkoulis_Lofi_Bonifati_Katsifodimos_2021}
\bibinfo{author}{Koutras, C.}, \bibinfo{author}{Siachamis, G.}, \bibinfo{author}{Ionescu, A.}, \bibinfo{author}{Psarakis, K.}, \bibinfo{author}{Brons, J.}, \bibinfo{author}{Fragkoulis, M.}, \bibinfo{author}{Lofi, C.}, \bibinfo{author}{Bonifati, A.}, \bibinfo{author}{Katsifodimos, A.}, \bibinfo{year}{2021}.
\newblock \bibinfo{title}{Valentine: Evaluating matching techniques for dataset discovery}, in: \bibinfo{booktitle}{2021 IEEE 37th International Conference on Data Engineering (ICDE)}.
\newblock \URLprefix \url{http://dx.doi.org/10.1109/icde51399.2021.00047}, \DOIprefix\doi{10.1109/icde51399.2021.00047}.
\bibitem[{Magnani et~al.(2005a)Magnani, Rizopoulos, McBrien and Montesi}]{Magnani_Rizopoulos_McBrien_Montesi_2005}
\bibinfo{author}{Magnani, M.}, \bibinfo{author}{Rizopoulos, N.}, \bibinfo{author}{McBrien, P.}, \bibinfo{author}{Montesi, D.}, \bibinfo{year}{2005}a.
\newblock \bibinfo{title}{Schema integration based on uncertain semantic mappings}.
\newblock \bibinfo{journal}{Lecture Notes in Computer Science,Lecture Notes in Computer Science} .
\bibitem[{Magnani et~al.(2005b)Magnani, Rizopoulos, Mc.Brien and Montesi}]{Magnani_Rizopoulos_Mc.Brien_Montesi_2005}
\bibinfo{author}{Magnani, M.}, \bibinfo{author}{Rizopoulos, N.}, \bibinfo{author}{Mc.Brien, P.}, \bibinfo{author}{Montesi, D.}, \bibinfo{year}{2005}b.
\newblock \bibinfo{title}{Schema Integration Based on Uncertain Semantic Mappings}.
\newblock p. \bibinfo{pages}{31–46}.
\newblock \DOIprefix\doi{10.1007/115683223}.
\bibitem[{Mork et~al.(2006)Mork, Rosenthal, Korb and Samuel}]{Mork_Rosenthal_Korb_Samuel_2006}
\bibinfo{author}{Mork, P.}, \bibinfo{author}{Rosenthal, A.}, \bibinfo{author}{Korb, J.}, \bibinfo{author}{Samuel, K.}, \bibinfo{year}{2006}.
\newblock \bibinfo{title}{Integration workbench: Integrating schema integration tools}, in: \bibinfo{booktitle}{22nd International Conference on Data Engineering Workshops (ICDEW’06)}.
\newblock \URLprefix \url{http://dx.doi.org/10.1109/icdew.2006.69}, \DOIprefix\doi{10.1109/icdew.2006.69}.
\bibitem[{Mudgal et~al.(2018)Mudgal, Li, Rekatsinas, Doan, Park, Krishnan, Deep, Arcaute and Raghavendra}]{Mudgal_Li_Rekatsinas_Doan_Park_Krishnan_Deep_Arcaute_Raghavendra_2018}
\bibinfo{author}{Mudgal, S.}, \bibinfo{author}{Li, H.}, \bibinfo{author}{Rekatsinas, T.}, \bibinfo{author}{Doan, A.}, \bibinfo{author}{Park, Y.}, \bibinfo{author}{Krishnan, G.}, \bibinfo{author}{Deep, R.}, \bibinfo{author}{Arcaute, E.}, \bibinfo{author}{Raghavendra, V.}, \bibinfo{year}{2018}.
\newblock \bibinfo{title}{Deep learning for entity matching}, in: \bibinfo{booktitle}{Proceedings of the 2018 International Conference on Management of Data}.
\newblock \URLprefix \url{http://dx.doi.org/10.1145/3183713.3196926}, \DOIprefix\doi{10.1145/3183713.3196926}.
\bibitem[{Pan et~al.(2023)Pan, Chan, Zou, Li, Basart, Woodside, Ng, Zhang, Emmons and Hendrycks}]{pan2023rewards}
\bibinfo{author}{Pan, A.}, \bibinfo{author}{Chan, J.S.}, \bibinfo{author}{Zou, A.}, \bibinfo{author}{Li, N.}, \bibinfo{author}{Basart, S.}, \bibinfo{author}{Woodside, T.}, \bibinfo{author}{Ng, J.}, \bibinfo{author}{Zhang, H.}, \bibinfo{author}{Emmons, S.}, \bibinfo{author}{Hendrycks, D.}, \bibinfo{year}{2023}.
\newblock \bibinfo{title}{Do the rewards justify the means? measuring trade-offs between rewards and ethical behavior in the machiavelli benchmark}.
\newblock \href{http://arxiv.org/abs/2304.03279}{\tt arXiv:2304.03279}.
\bibitem[{Popa et~al.(2002)Popa, Velegrakis, Miller, Hernández and Fagin}]{Popa_Velegrakis_Miller_Hernández_Fagin_2002}
\bibinfo{author}{Popa, L.}, \bibinfo{author}{Velegrakis, Y.}, \bibinfo{author}{Miller, R.J.}, \bibinfo{author}{Hernández, M.A.}, \bibinfo{author}{Fagin, R.}, \bibinfo{year}{2002}.
\newblock \bibinfo{title}{Translating Web Data}.
\newblock p. \bibinfo{pages}{598–609}.
\newblock \URLprefix \url{http://dx.doi.org/10.1016/b978-155860869-6/50059-7}, \DOIprefix\doi{10.1016/b978-155860869-6/50059-7}.
\bibitem[{Qiang et~al.(2024)Qiang, Wang and Taylor}]{Qiang_Wang_Taylor_2024}
\bibinfo{author}{Qiang, Z.}, \bibinfo{author}{Wang, W.}, \bibinfo{author}{Taylor, K.}, \bibinfo{year}{2024}.
\newblock \bibinfo{title}{Agent-om: Leveraging llm agents for ontology matching} .
\bibitem[{Radwan et~al.(2009)Radwan, Popa, Stanoi and Younis}]{radwan2009top}
\bibinfo{author}{Radwan, A.}, \bibinfo{author}{Popa, L.}, \bibinfo{author}{Stanoi, I.R.}, \bibinfo{author}{Younis, A.}, \bibinfo{year}{2009}.
\newblock \bibinfo{title}{Top-k generation of integrated schemas based on directed and weighted correspondences}, in: \bibinfo{booktitle}{Proceedings of the 2009 ACM SIGMOD International Conference on Management of data}, pp. \bibinfo{pages}{641--654}.
\bibitem[{Rahm and Bernstein(2001)}]{rahm2001survey}
\bibinfo{author}{Rahm, E.}, \bibinfo{author}{Bernstein, P.A.}, \bibinfo{year}{2001}.
\newblock \bibinfo{title}{A survey of approaches to automatic schema matching}.
\newblock \bibinfo{journal}{the VLDB Journal} \bibinfo{volume}{10}, \bibinfo{pages}{334--350}.
\bibitem[{Roitman et~al.(2008)Roitman, Gal and Domshlak}]{2008Providing}
\bibinfo{author}{Roitman, H.}, \bibinfo{author}{Gal, A.}, \bibinfo{author}{Domshlak, C.}, \bibinfo{year}{2008}.
\newblock \bibinfo{title}{Providing Top-K Alternative Schema Matchings with Onto Matcher}.
\newblock \bibinfo{publisher}{Conceptual Modeling - ER 2008}.
\bibitem[{Sheetrit et~al.(2024)Sheetrit, Brief, Mishaeli and Elisha}]{sheetrit2024rematch}
\bibinfo{author}{Sheetrit, E.}, \bibinfo{author}{Brief, M.}, \bibinfo{author}{Mishaeli, M.}, \bibinfo{author}{Elisha, O.}, \bibinfo{year}{2024}.
\newblock \bibinfo{title}{Rematch: Retrieval enhanced schema matching with llms}.
\newblock \bibinfo{journal}{arXiv preprint arXiv:2403.01567} .
\bibitem[{Shvaiko and Euzenat(2005)}]{Shvaiko_Euzenat_2005}
\bibinfo{author}{Shvaiko, P.}, \bibinfo{author}{Euzenat, J.}, \bibinfo{year}{2005}.
\newblock \bibinfo{title}{A survey of schema-based matching approaches}.
\newblock p. \bibinfo{pages}{146–171}.
\newblock \DOIprefix\doi{10.1007/116034125}.
\bibitem[{T\"ornberg(2023)}]{Tornberg_2023}
\bibinfo{author}{T\"ornberg, P.}, \bibinfo{year}{2023}.
\newblock \bibinfo{title}{Chatgpt-4 outperforms experts and crowd workers in annotating political twitter messages with zero-shot learning} .
\bibitem[{Tu et~al.()Tu, Fan, Tang, Wang, Du and Wang}]{Tu_Fan_Tang_Wang_Du_Wang}
\bibinfo{author}{Tu, J.}, \bibinfo{author}{Fan, J.}, \bibinfo{author}{Tang, N.}, \bibinfo{author}{Wang, C.}, \bibinfo{author}{Du, X.}, \bibinfo{author}{Wang, P.}, .
\newblock \bibinfo{title}{Unicorn: A unified multi-tasking model for supporting matching tasks in data integration} .
\bibitem[{Zhang et~al.(2020)Zhang, Chen, Jagadish, Zhang and Tong}]{zhang2020reducing}
\bibinfo{author}{Zhang, C.J.}, \bibinfo{author}{Chen, L.}, \bibinfo{author}{Jagadish, H.}, \bibinfo{author}{Zhang, M.}, \bibinfo{author}{Tong, Y.}, \bibinfo{year}{2020}.
\newblock \bibinfo{title}{Reducing uncertainty of schema matching via crowdsourcing with accuracy rates}.
\newblock \bibinfo{journal}{IEEE Transactions on Knowledge \& Data Engineering} \bibinfo{volume}{32}, \bibinfo{pages}{135--151}.
\bibitem[{Zhang et~al.()Zhang, Floratou, Cahoon, Krishnan, Müller, Banda, Psallidas and Patel}]{Zhang_Floratou_Cahoon_Krishnan_Müller_Banda_Psallidas_Patel}
\bibinfo{author}{Zhang, Y.}, \bibinfo{author}{Floratou, A.}, \bibinfo{author}{Cahoon, J.}, \bibinfo{author}{Krishnan, S.}, \bibinfo{author}{Müller, A.}, \bibinfo{author}{Banda, D.}, \bibinfo{author}{Psallidas, F.}, \bibinfo{author}{Patel, J.}, .
\newblock \bibinfo{title}{Schema matching using pre-trained language models} .

\end{thebibliography}






\end{document}